\documentclass{aastex63}
\usepackage{epsfig,amsmath,amsfonts,amssymb,graphicx,subfigure,enumitem,color}
\usepackage{amssymb}
\usepackage{amsmath}
\usepackage[varg]{txfonts}
\usepackage{natbib}
\usepackage{url}             
\usepackage{graphicx}
\usepackage{float}
\newcommand{\beq}{\begin{equation}}
\newcommand{\eeq}{\end{equation}}
\newcommand{\beqa}{\begin{eqnarray}}
\newcommand{\eeqa}{\end{eqnarray}}
\newcommand{\beqann}{\begin{eqnarray*}}
\newcommand{\eeqann}{\end{eqnarray*}}
\bibpunct{(}{)}{;}{a}{}{,} 

\usepackage{textcomp}
\newcommand{\foo}{\rlap{+}{\texttimes}}
\shortauthors{Mishra et al.}
\shorttitle{Kelvin-Helmholtz Instability}
\begin{document}

\title{ Evolution of Kelvin-Helmholtz Instability in the Fan-Spine Topology}
\correspondingauthor{Sudheer K. Mishra}
\email{sudheerkm.rs.phy16@itbhu.ac.in}
\author[0000-0003-2129-5728]{Sudheer K.~Mishra}
\affil{Indian Institute of Astrophysics, Sarjapur Main Road, 2$^{nd}$ Block, Koramangala, Bangalore-560034, India}
\author{Balveer Singh}
\affil{Department of Physics, Indian Institute of Technology (BHU), Varanasi-221005, India.}
\author{A.K.~Srivastava}
\affil{Department of Physics, Indian Institute of Technology (BHU), Varanasi-221005, India.}
\author{Pradeep Kayshap}
\affil{VIT Bhopal, Kothari Kalan, Sehore, Madhya-Pradesh 466114, India}
\author{B.N.~Dwivedi}
\affil{Department of Physics, Indian Institute of Technology (BHU), Varanasi-221005, India.}

\bigskip

\begin{abstract}
We use multiwavelength imaging observations from the Atmospheric Imaging Assembly (AIA) onboard the \textit{ Solar Dynamics Observatory (SDO)} to study the evolution of Kelvin-Helmholtz (K-H) instability in a fan-spine magnetic field configuration. This magnetic topology exists near an active region AR12297 and is rooted in a nearby sunspot. In this magnetic configuration, two layers of cool plasma flow in parallel and interact with each other inside an elongated spine. The slower plasma flow (5 $km s^{-1}$) is the reflected stream along the spine’s field lines from the top, which interacts with the impulsive plasma upflows (114–144 $km s^{-1}$) from below. This process generates a shear motion and subsequent evolution of the K--H instability. The amplitude and characteristic wavelength of the K-H unstable vortices increase, satisfying the criterion of the fastest growing mode of this instability. We also describe that the velocity difference between two layers and velocity of K-H unstable vortices are greater than the Alfv\'en speed in the second denser layer, which also satisfies the criterion of the growth of K-H instability. In the presence of the magnetic field and sheared counter streaming plasma as observed in the fan-spine topology, we estimate the parametric constant, $\Lambda\ge$1, that confirms the dominance of velocity shear and the evolution of the linear phase of the K-H instability. This observation indicates that in the presence of complex magnetic field structuring and flows, the fan-spine configuration may evolve into rapid heating, while the connectivity changes due to the fragmentation via the K-H instability.
\end{abstract}

\section{Introduction}
Magnetic reconnection is a novel physical plasma process in which the complex magnetic structures reorganize to form simpler magnetic field configurations. In this process, the accumulated magnetic energy is released in the form of kinetic energy, heat, radiation, etc., in the solar atmosphere \citep{2007mare.book.....P, 2010RvMP...82..603Y, 2014masu.book.....P}. It also serves as one of the fundamental mechanisms in the solar atmosphere to trigger various types of solar eruptions and can also be one of the major candidates of the coronal heating \citep{1988ApJ...330..474P, 1994Natur.371..495M, 1995Natur.375...42Y, 1997Natur.386..811I, 2004ApJ...605..911C, 2011LRSP....8....6S, 2015Natur.523..437S, 2016NatPh..12..847L, 2016NatCo...711837X, 2019ApJ...887..137S, 2020SoPh..295..167M, 2021arXiv210706940S}. In recent years, high-resolution multi-instrument observations, as well as simulations, are focused on locating the 3D magnetic null points and understanding reconnection and dynamical processes there \citep{2011AdSpR..47.1508P, 2011SSRv..158..205M, 2009PhPl...16l2101P}. The null point is considered a location where the magnetic topology sharply changes due to the magnetic reconnection, and also the maximum energy release may also occur. Two topological magnetic structures (fan and spine) pass through the null point \citep{1996RSPSA.354.2951P}. The fan surface is defined as a plane that evolves from the null point. The spine is defined as the bundle of field lines that approach or move away from the null point. The fan-spine topology is favourable for the triggering of solar flares \citep{1990ApJ...350, 1996RSPSA.354.2951P, 2013PhPl...20c2117W, 2017Natur.544..452W, 2018NatSR...8.8136L, 2019ApJ...885L..11S}. The fan-spine reconnection over the null point is also responsible for triggering the circular ribbon in the lower solar atmosphere \citep{2009ApJ...700..559M, 2012A&A...547A..52R, 2016ApJ...827...27Z, 2019ApJ...874L..33M, 2020A&A...636L..11Z}. In the fan-spine topology, when the spine is open in the coronal region, it causes recurring jets and CMEs. These extended spines also expel plasma material from the null point to another end of the spine and create remote brightening \citep{2013NatPh...9..489S, 2017ApJ...847..124H, 2018ApJ...857..115Y, 2020ApJ...898..101Y}. Both the fan-spine topology and torsional fan-spine topology possess shear in the magnetic fields and velocity streamlines. The sheared magnetic fields and velocity streamlines layers may become unstable and cause the fragmentation of this magnetic configuration. This fragmentation is responsible for the evolution of the tearing mode instability or the Kelvin-Helmholtz instability, which depends on the magnetic and velocity shear dominance. \citet{2013PhPl...20c2117W} have simulated the 3D null point in the fan-spine configuration and found that the Kelvin-Helmholtz instability is evolved in a current-vortex sheet.  \\

The solar atmosphere possesses several small to large-scale eruptions (e.g., spicules, jets, prominences, CMEs, flares, etc.), MHD waves, instabilities, energetics, etc., at the different spatio-temporal scales. The magnetohydrodynamic instabilities play a crucial role in the evolution and eruption of the small to large-scale magnetic structures and plasma ejecta \citep{2015A&A...574A..55Z, 2019ApJ...874...57M, 2018ApJ...856...86M, 2018ApJ...856...44A, 2020ApJ...893L..46Z}. As stated above, the MHD instabilities as well as gravity-driven instabilities (e.g., magnetic Rayleigh-Taylor instability, hybrid K-H,\--\,R-T instability, etc.) and shear flow instability (K-H) develop in the solar corona at different spatio-temporal scales and trigger small to large-scale eruptions. The K-H instability is a shear flow-driven instability that evolves at the interface of two fluids when they undergo the differential sheared velocity at the interface. This instability was initially discussed by Kelvin (1871) and Helmholtz (1868). The K-H instability develops at the interface of two fluids if there is a velocity gradient ($\Delta \vec{u}$=$\vec{u}_{1}$-$\vec{u}_{2}$). The flow velocity of the upper fluid ($\vec{u}_{1}$) needs to exceed the Alfv\'en velocity ($\vec{u}_{A2}$) of the second fluid at the interface to produce a K-H instability. This condition was firstly investigated by \citet{1978SoPh...58...57P}. The Kelvin-Helmholtz instability evolves in two different manners. The first physical mechanism is an evolution of the surface mode instability in which the differential velocity shear triggers the formation of multiple vortices at the interface of two fluids. It evolves in a variety of the astrophysical systems, in the atmosphere of the Earth and planets, gaseous, fluids, and plasma \citep{1993ApJ...407..588M, 1997ApJ...483..262V, 2010Natur.466..947B, 2014SSRv..184....1J}. It is one of the major causes of the evolution of turbulence in astrophysical plasmas. The solar atmosphere possesses magnetized plasma. There are several theoretical studies of Kelvin-Helmholtz instability as a boundary layer problem in the magnetized plasma \citep{1983A&A...117..220H, 1983A&A...126..209R, 2001A&A...368.1083A, 2004A&A...420..737K, 2010ApJ...712..875S, 2012ApJ...749..163S, 2010A&A...516A..84Z}. In the presence of the magnetic field, the magnetic tension component of the Lorentz force and compressibility act as stabilizing effects to reduce the instability. The linear growth of the Kelvin-Helmholtz instability was analytically calculated for the sheared flows in an idealized fluid \citep{1961hhs..book.....C}. \\

The evolution of K-H instability is observed on the different spatio-temporal scales and possesses significant dynamics (e.g., vortex, turbulence, fingers, plumes, etc.) in the solar atmosphere. The gravity-driven magnetic Rayleigh-Taylor and shear-flow-driven K-H instabilities are responsible for evaporating the cool prominence plasma in the hot solar corona. This instability develops at the prominence-corona interface in the form of the ripples, plumes, and vortex-like structures \citep{2008ApJ...676L..89B, 2010ApJ...716.1288B,2010SoPh..267...75R, 2017ApJ...850...60B, 2018ApJ...857..115Y, 2018ApJ...864L..10H, 2018ApJ...856...86M, 2019ApJ...874...57M}. The first observational evidence of K-H instability in EUV waveband is reported by \citet{2011ApJ...734L..11O}. They have found that the vortex-shaped structure is evident in between the erupting and non-erupting regions of the solar atmosphere. These vortex-like structures also develop at the interface of the eruptive CME and ambient corona \citep{2011ApJ...729L...8F, 2013ApJ...767..170F, 2013ApJ...766L..12M}. The K-H unstable vortices also develop in the rotating coronal jets, blow-out jets, and polar coronal jets \citep{2018NatSR...8.8136L, 2015AdSpR..56.2727Z, 2016Ap&SS.361...51Z, 2019FrASS...6...33Z, 2019SoPh..294...68S}. As discussed above, the observation of K-H instability is well studied in the cool and dense prominences, jets, at the interface of the CME and ambient corona, and near the flank of CME \citep{2010SoPh..267...75R, 2011ApJ...729L...8F, 2013ApJ...767..170F, 2013ApJ...766L..12M, 2017ApJ...850...60B, 2018ApJ...857..115Y, 2018ApJ...864L..10H, 2018NatSR...8.8136L, 2019ApJ...874...57M, 2019SoPh..294...68S}. However, the evolution of K-H instability in the hot region of the corona above dynamical fan-spine topology is not reported yet.
\\ 

In the present paper, we provide a novel observational study of the evolution of the Kelvin-Helmholtz (K-H) instability in a hot fan-spine topology. The K-H unstable vortices are associated with the cool plasma and appear inside the hot spine loops. Using multi-thermal EUV wavebands of the \textit{SDO}/AIA, we find that multi-thermal layers of the cool plasma interact with each other and cause the onset of K-H instability in an elongated spine. The hot spine topology is present as an ambient at 8--10 MK in the solar corona. The two layers of the cool plasma flow interact with each other layer inside a hot elongated spine and cause the onset of the K-H instability. The observation and data analyses techniques are discussed in Section~2. Section~3 outlines the observational result. The discussions and conclusion are presented in Section~4.
\begin{figure*}
\includegraphics[scale=1.0,angle=90,width=18.0cm,height=18.0cm,keepaspectratio]{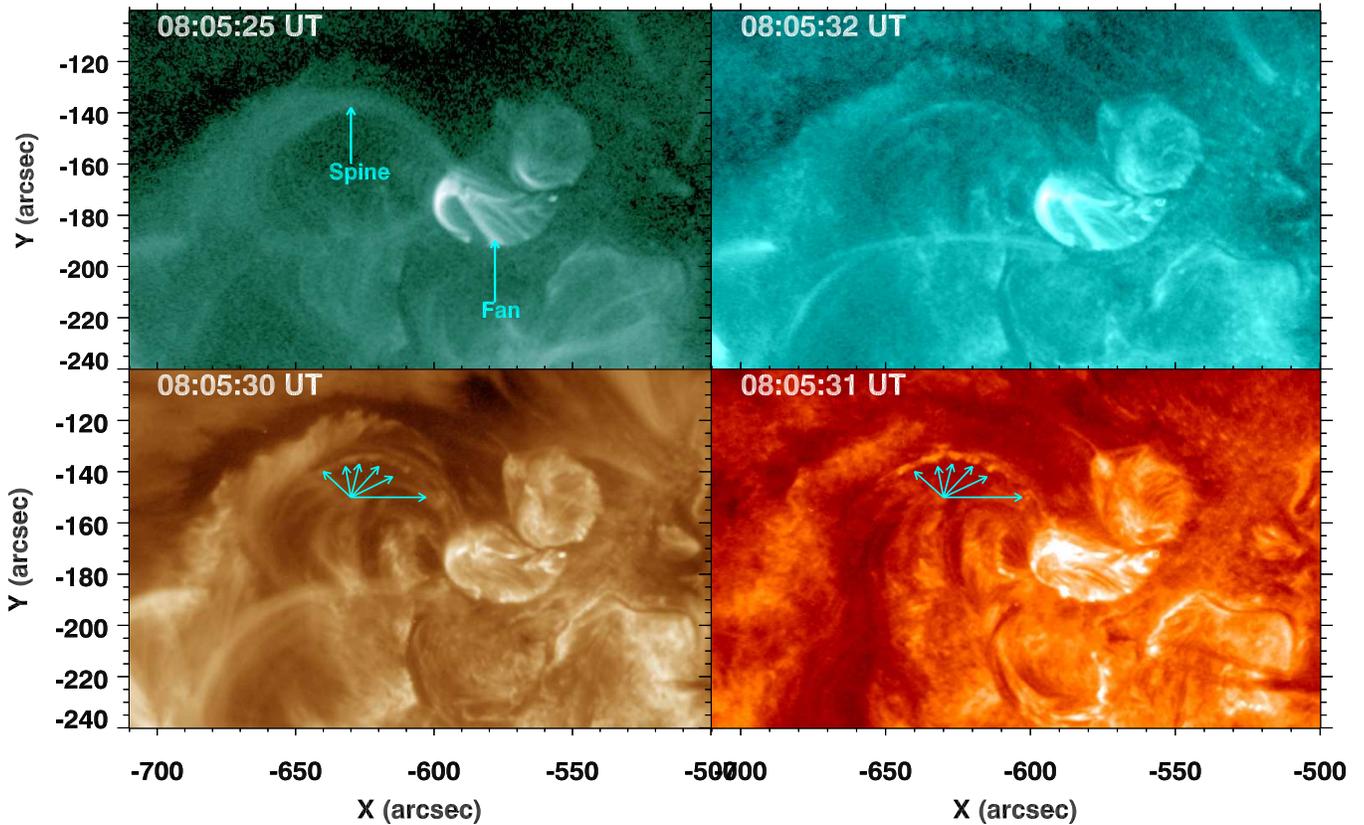}
\caption{Multi-temperature view of the fan-spine topology in hot and cool plasma as displayed in the \textit{SDO}/AIA 94 {\AA}, 131 {\AA}, 193 {\AA}, 304 {\AA} filters at 08:05 UT on 2015 March 10. A dome-shaped fan-spine topology is visible in hot wavebands (e.g., 94 {\AA}, 131 {\AA}), while the K-H unstable vortices and two different layers appeared in the cool wavebands (e.g., 304 {\AA}) and at typical coronal temperature e.g., 171 {\AA}, 193 {\AA}). Multithermal K-H unstable vortices have indicated by the cyan-color arrows.}
\end{figure*}
\section{Observational Data and Analyses} 
We use high-resolution and high-cadence data obtained from the {\textit Atmospheric Imaging Assembly} \citep{2012SoPh..275...17L} onboard \textit{Solar Dynamics Observatory} \citep{2012SoPh..275....3P}. The AIA telescope possesses seven EUV channels (e.g., 94,  131, 171, 193, 211, 304, 335) {\AA}, two UV channels (1600, 1700) {\AA}, and a visible channel (4500) {\AA} to observe the different layers of the Sun's atmosphere covering from the photosphere to the lower corona. The AIA/EUV channels provide 4096$\times$ 4096-pixel full-disk images of the Sun with a spatial resolution of 1.5 arcsec, with a pixel size of 0.6 arcsec. The cadence of the temporal image data is 12 seconds. The aia\_prep.pro routine available in the SSW Solarsoft library is used for the basic analysis and calibration of the AIA data.\\

The properties of the multithermal plasma associated with the evolution of the Kelvin-Helmholtz instability are understood by evaluating the differential emission measure (DEM). We use the method of \citep{2015ApJ...807..143C} to measure the DEM in the region-of-
interest where the evolution of the Kelvin-Helmholtz unstable cool plasma and surrounding hot loops are present. This method map the DEM with the different temperature using six AIA filters, i.e, 94 {\AA}, 131 {\AA}, 171 {\AA}, 193 {\AA}, 211 {\AA}, 335 {\AA} in the framework of the sparse inversion method. The sparse inversion method takes some data points with an underdetermined linear system. The sparse inversion code available in Solarsoft and written in IDL is used to extract the differential emission measure (DEM) from a few data points of AIA. This method adopts the concept of sparsity and uses a simplex function to derive the solution from the optically thin \textit{SDO}/AIA filters by minimizing the total emission measure (EM). The sparse inversion technique provides only the positive solution in the range of max(0, I-tol) to min(0, I+tol), where I is reconstructed intensity and tol is tolerance in the reconstructed intensities. We have chosen the temperature between log $T$(K)=5.0--7.5 with 25 temperature bins at log $T$(K)=0.1 intervals to deduce the DEM. Using the DEM analysis, we identify the cool plasma visible in the temperature range log $T$(K)=5.0--6.3. The major part of the Kelvin-Helmholtz unstable vortices is associated with the cool plasma observed by the cool coronal EUV filters, e.g., 304, 171, 193, 211 {\AA} of AIA channels. The hot loops are mostly visible in the temperature range of log $T$(k)=6.8--7.2, which corresponds to AIA 94 {\AA} and AIA 131 {\AA} filters. To estimate the density, we use the estimated total emissions coming from different temperatures given as $n$=$\sqrt{\frac{EM}{l}}$, where $n$ is the number density, $EM$ is the total emission coming from the different temperature bins, and $l$ is the depth of K-H unstable vortices. To deduce the density within the K-H unstable vortex and overlying layer, we assume that the width of the vortices is equal to the depth. Using the aforementioned observational data and various analysis techniques, we observe a fan-spine configuration consisting of K-H instability in the elongated spine. The detailed observational result and their physical implications are presented in the forthcoming section.\\

\section{Observational Results}
We observe a fan-spine configuration using multi-wavelength imaging data of \textit{SDO}/AIA on 2015 March 10 (Figure~1). This typical fan-spine configuration situates near the AR 12297 without any associated remote brightening. The spine loops are inherent and mostly appear in the hot AIA channels. Multi-thermal plasma flows inside the elongated spine along the magnetic field lines (Figure~1). We use AIA 304 {\AA} images with reverse color contrast showing the spatio-temporal variation of flows (L1, L2; Figure~2) to understand the two-stream flow inside the spine. Initially, at 07:48 UT, a thin plasma layer (L1; Figures 2–3) started to lift along the elongated length of the spine. As this layer reaches its maximum height inside the hot spine, plasma deceleration starts after $\approx$07:57 UT in the upper layer (Figures~2–3). Another layer of plasma flows (L2; Figures~2--3) initiates at $\approx$07:58 UT. These two flows (L1, L2; Figures 2–3) pass along the maximum height of the spine-related field lines. At a maximum height of the spine, these two flows have a significant velocity difference that leads to the onset of K-H instability. Due to the substantial shear velocity, the smooth boundary turned into a sawtooth boundary, as shown in Figure~3 (cyan dotted lines). The smooth boundary exhibits a sawtooth pattern, indicating the onset of the K-H instability in the lower plasma flow (L2; Figure~3). As the K-H instability developed in the lower layer of flow (L2), multiple vortex-like structures have formed within 90 seconds (Figure~3). We observe that the sheared magnetic field may have reconnected various times in the fan plane and causes the launch of multiple plasmas flows along the elongated spines. The appearance of hot plasma and localized brightening observed in the hot EUV channel of {\textit{SDO}/AIA may arise due to these multiple reconnections in the sheared fan plane (Figure~1). To mimic the physical scenario and magnetic field structuring in the lower corona, we draw a schematic to discuss the fan-spine configuration, different layers of plasma flow, and development of the K-H unstable vortices in the elongated hot spine (Figure~4). It describes the dynamical evolution and overall magnetic configuration containing the fan-spine topology, two streams of flows (L1, L2), interaction of two fluids inside the spine, and formation of vortices-like structures due to the evolution of K-H instability in the solar corona (green and yellow lines; Figure~4). The yellow lines represent the fan surface, and
overlying spine-associated field lines are presented as green lines. The long spine-associated field lines connect
remotely in the solar corona without creating any remote brightening (Figures~1, 4). Two layers of plasma flows propagate along the
spine-associated magnetic field (blue and brown dotted lines). At the maximum height, these two flows interact and are responsible for the onset of K-H unstable vortices (Figures~3--4). Two plasma blobs shoot up along this layer with the velocity of $\approx$58 km s$^{-1}$ and 66 km s$^{-1}$ (Figure~5). As this layer reaches its maximum height inside the hot spine, plasma deceleration starts after $\approx$07:57 UT in the upper layer (Figures~2–5). The second layer of plasma flow (L2) is initiated to lift at $\approx$07:58 after the deceleration of the first layer begins. These two layers propagate in the parallel direction with a velocity of $\approx$5km s$^{-1}$ and 114–144 km s$^{-1}$, respectively (Figure~5). Therefore, the estimated velocity-difference or shearing velocity ($\Delta \vec{u}$=$\vec{u}_{1}$-$\vec{u}_{2}$) near the interface of two fluids is lying between 109–139 km s$^{-1}$. The K-H instability develops in the lower layer of plasma flow due to the strong velocity shear ($\approx$109--139 km s$^{-1}$) between these two flows (Figure~5). 
\\

The K-H instability is a fundamental instability that may evolve in hydrodynamic (HD) and magnetized plasma. It develops due to the velocity differences in the two fluids, resulting in the formation of vortices at the interface of the two fluids. In the magnetized plasma, the magnetic field is responsible for the suppression of the instability. The tension component of the Lorentz force suppresses the growth of the vortices. As discussed above, the K-H instability evolved at the two-fluid interface. The lower dense fluid lifts the higher dense fluid, and higher dense fluid pushes down the low denser fluid at the interface, so overall, a gain of potential energy occurs. This sharp density gradient and shear flow are responsible for forming the vortices at the two fluids interface. As a result, K-H instability evolves at the interface and rolls up of fluid begins. For an incompressible fluid, if two fluids have different densities (the heavier fluid being above, the lighter fluid), different velocities, and lying at the same horizontal magnetic interface, the dispersion relation of frequency ($\omega$) and wave vector ($\vec{k}$) is given as \citep{2017ApJ...850...60B}, 
\begin{equation}
\omega=(\alpha_{1}\vec{k}.\vec{u_{1}}+\alpha_{2}\vec{k}.\vec{u_{2}})\pm i\Big[\Big\{gk(\alpha_{1}-\alpha_{2})- \frac{(\vec{k}.\vec{B_{1}})^{2}+(\vec{k}.\vec{B_{2}})^{2}}{2\pi(\rho_{1}+\rho_{2})}\Big\}+\alpha_{1}\alpha_{2}(\vec{k}.\Delta \vec{u})^{2}\Big]^\frac{1}{2}
\end{equation}
where $\alpha_{1}$=$\frac{\rho_{1}}{(\rho_{1}+\rho_{2})}$, $\alpha_{2}$=$\frac{\rho_{2}}{(\rho_{1}+\rho_{2})}$ is a dimensionless ratio of density, g is gravity, $\vec{k}$=($ 2\pi/\lambda$) is wave vector, $\lambda$ is the characteristic wavelength of the K-H instability, which defines the separation between two consecutive vortices, $\vec{u}_{1}$ and $\vec{u}_{2}$ are the velocity of plasma in the upper and lower layer, $\vec{B}_{1}$ and $\vec{B}_{2}$ are magnetic field in the lower and upper layer, and $\omega$ is frequency. 
Examining the dispersion relation (Equation~1) shows that it has two components: a real part Re($\omega$), which leads to the stable oscillations or surface wave solutions and the imaginary part Im($\omega$)=$\gamma$ that defines the exponential growth rate of the K-H and RT instability \citep{2017ApJ...850...60B, 2018RvMPP...2....1H}. Therefore, the growth rate of the instability ($\gamma$) is given by,
\begin{figure*}
\includegraphics[scale=1.0,angle=0,width=18.0cm,height=18.0cm,keepaspectratio]{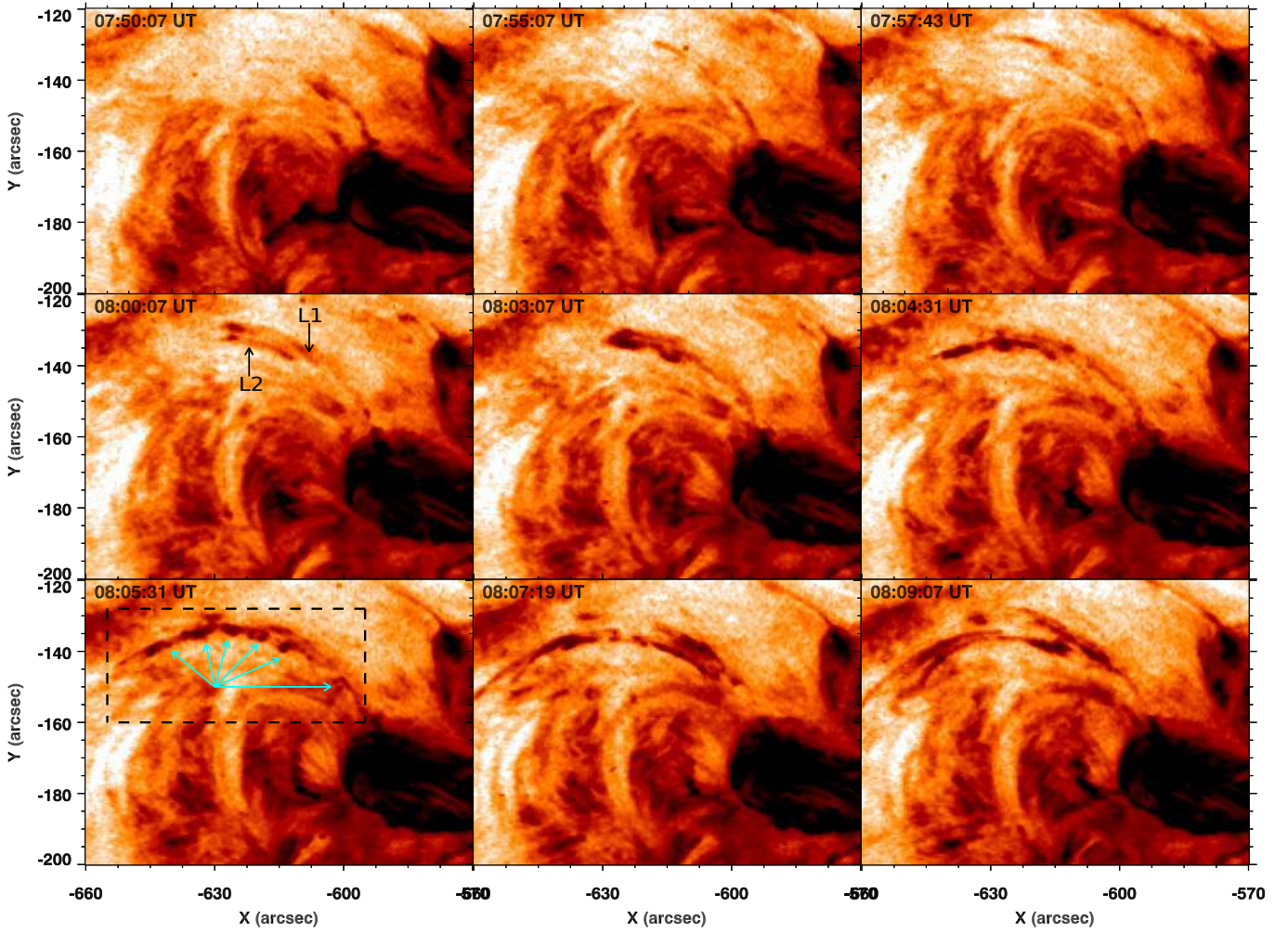}
\caption{The sequence of images of \textit{SDO}/AIA 304 {\AA} shows the two layers of plasma flow and the onset of K-H instability. Two layers of plasma flows are indicated by the L1 and L2. Different cyan-colored arrows indicate the K-H unstable vortices.}
\end{figure*}
\begin{equation}
\gamma^{2}=gk{\big(\alpha_{1}-\alpha_{2}\big)}-\Big[\frac{(\vec{k}.\vec{B_{1}})^{2}+(\vec{k}.\vec{B_{2}})^{2}}{\mu_{0}(\rho_{1}+\rho_{2})}-\alpha_{1}\alpha_{2}(\vec{k}.\Delta \vec{u})^{2}\Big]
\end{equation}
When $\rho_{1}$ $>$ $\rho_{2}$ and $\vec{u}_{1}$ $\neq$ $\vec{u}_{2}$, and in the absence of the magnetic field, the hydrodynamic K-H instability appears. Therefore, in the absence of the magnetic field or if the magnetic field is perpendicular to the interface of the two fluids, the condition for the onset of hydrodynamic K-H instability in every other direction \citep{1961hhs..book.....C, 1998pfp..book.....C, 2021JGRA..12629097S} is,
\begin{figure*}
\includegraphics[scale=1.0,angle=0,width=18.0cm,height=18.0cm,keepaspectratio]{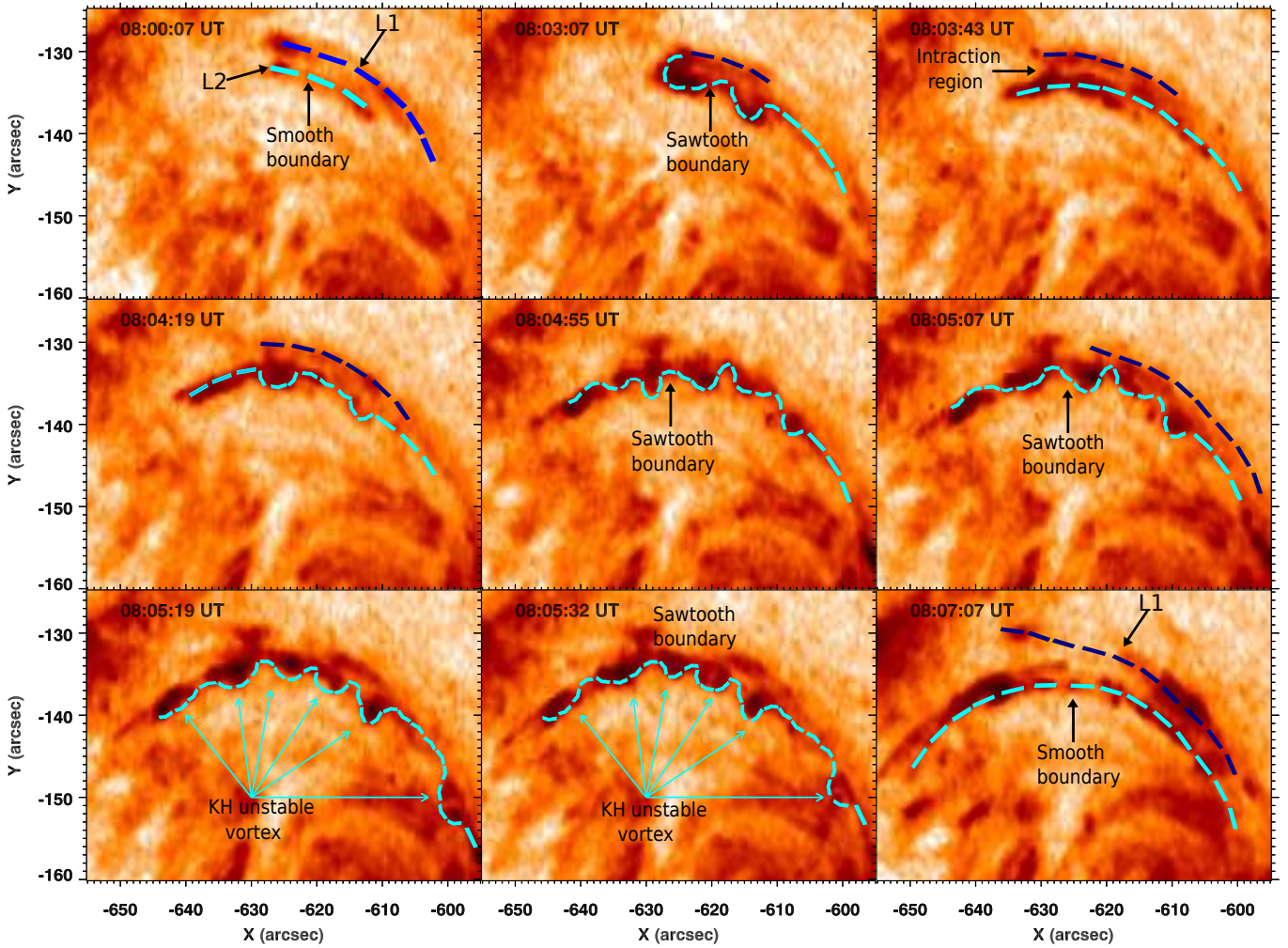}
\caption{The region of interest (ROI; black box in Figure 2) is shown by \textit{SDO}/AIA 304 {\AA} filter. Two plasma flow layers are indicated by ''L1'' and ''L2'' as indicated by the blue and cyan color dotted lines. Initially, the boundary of these two flows is smooth; however, when the interaction of two layers starts, it becomes a sawtooth pattern. Cyan-colored arrows indicate that the K-H unstable vortex develops in the lower flow layer (L2). Later, the K-H unstable vortices merged within each other, and again boundary became smooth.}
\end{figure*}
\begin{equation}
\alpha_{1}\alpha_{2}(\vec{k}.\Delta \vec{u})^{2}>gk(\alpha_{1}-\alpha_{2})
\end{equation}
The growth rate of the instability after substituting the value of $\alpha_{1,2}$ in the presence of the magntic field and grouping terms differently gives \citep{2017ApJ...850...60B, 2018RvMPP...2....1H}, 
\begin{equation}
\gamma^{2}=gkA-\Big[\frac{(\vec{k}.\vec{B_{1}})^{2}+(\vec{k}.\vec{B_{2}})^{2}}{\mu_{0}(\rho_{1}+\rho_{2})}-\frac{{\rho_{1}}{\rho_{2}}}{(\rho_{1}+\rho_{2})^{2}}(\vec{k}.\Delta \vec{u})^{2}\Big]
\end{equation}

Where A is Atwood number defined as A=$\frac{\rho_{1}-\rho_{2}}{\rho_{1}+\rho_{2}}$, and $\gamma$ is the growth rate of the instability. The first term in Equation~2 is related to the gravity-driven (density gradient) instability, which leads to the hydrodynamic Rayleigh-Taylor instability at the interface of two fluids. The other two terms deal with the balance between the magnetic tension force and shear flow and the density inversion interface. Therefore the gravity (also density gradient) and shear flow are responsible for the instability, while the magnetic tension component suppresses the instability. The above equation (4) contains criteria for magnetic RT, K-H, and the hybrid R-T and K-H instabilities \citep{2017ApJ...850...60B, 2018RvMPP...2....1H, 2019ApJ...874...57M}. \\
\begin{figure*}
\includegraphics[scale=1.0,angle=0,width=18.0cm,height=18.0cm,keepaspectratio]{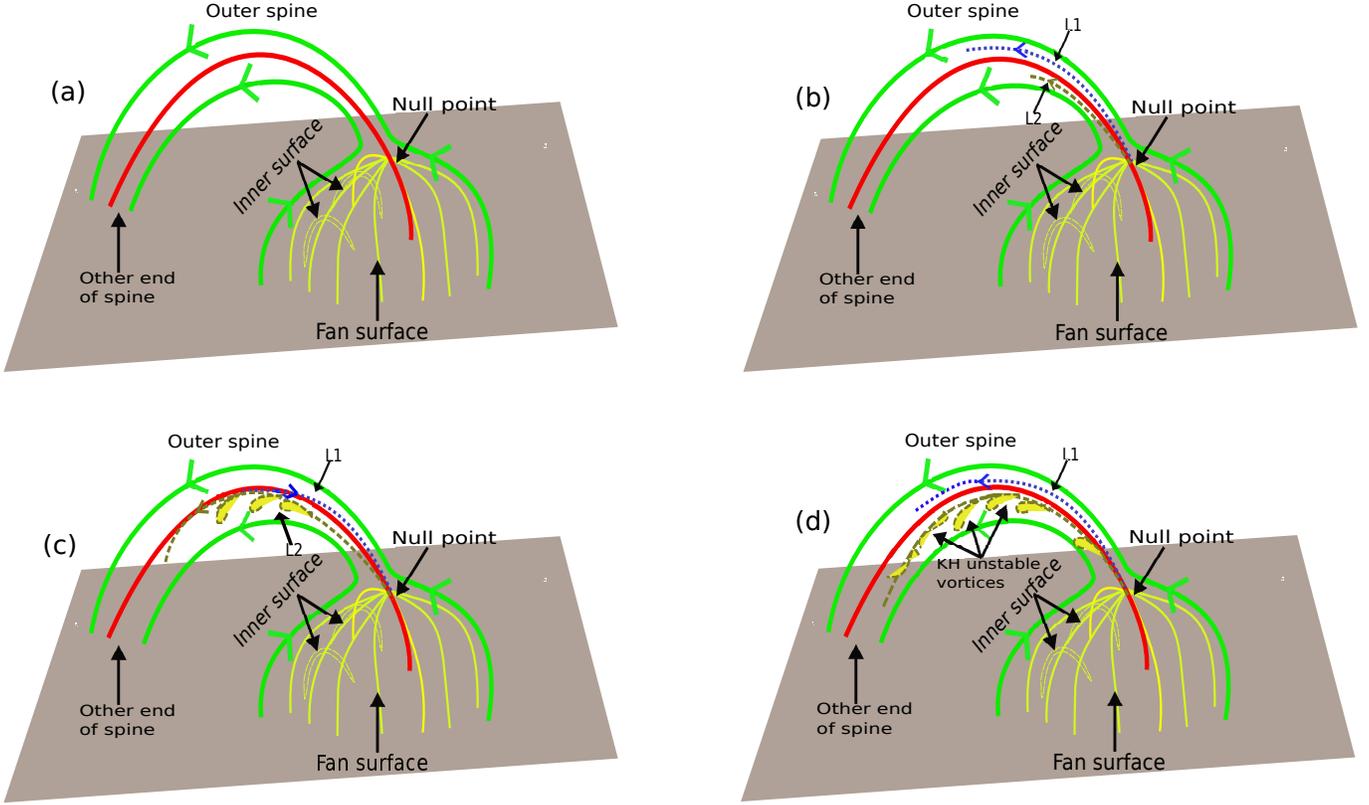}
\caption{A schematic shows the dynamical evolution of multi-layer plasma flow inside the ambient hot spine and the overall magnetic configuration of the fan-spine topology. The green color lines indicate the outer spine associated magnetic field. The yellow color lines collectively define the fan surface. The magnetic field lines (yellow) approach the null point. The null point is lying above the fan surface and has indicated by the red line. A thin layer of cool plasma (layer L1) lifts and flows along the elongated length of the spine. This layer reaches the highest height, and after that, the second layer of plasma flow starts (L2; brown dotted line lying inside the green outer spines). At the highest height of the spine, the two-layer of plasma flow propagates in opposite directions. A layer of invisible material separates the two layers of plasma (blue and brown). At this invisible surface, multiple K-H unstable vortices appeared due to the velocity difference.}
\end{figure*}
We adopt the sparse inversion code to estimate the mass densities within the higher dense K-H unstable vortices and in the upper layer (L1) \citep{2015ApJ...807..143C}. We select a vortex structure within 10$\times$10 arcsec box region as shown in the box region of Figure~6. The total emission (EM) coming from the bright K-H unstable vortex region and overlying region are 2.5$\times$10$^{28}$ cm$^{-5}$ and 1.6$\times$10$^{28}$ cm$^{-5}$ respectively. The plasma density \Big($n$=$\sqrt{\frac{EM}{l}}$\Big) associated with these two regions are 5.8$\times$10$^{9}$ cm$^{-3}$ and 4.7$\times$10$^{9}$ cm$^{-3}$ respectively. We assume that the maximum depth and width of the vortex are the same ($\approx$7.25 Mm, Figure~6). The cool solar plasma possesses 90\% hydrogen and 10\% helium \citep{2010SSRv..151..243L, 2011ApJ...727...25G}. The estimated mass densities for K-H unstable vortices and overlying plasma flow is 7.6$\times$ 10$^{-15}$ and 6.1$\times$ 10$^{-15}$ g cm$^{-3}$ respectively (Figure~6). In the present observation, we found that there is a not significant difference in the two layers of the densities ($\rho_{1}, \rho_{2};$ Figure~6). The Rayleigh-Taylor instability term (gravity term) is maximum when the Atwood number is high. In the present observation, the estimated value of the Atwood number is $\approx$0.1. Therefore, the gravity-driven instability term ignores as compared to the shear flow term that has scaled with the $\frac{\rho_{1}}{\rho_{2}}=1.25$ \citep{2017ApJ...850...60B, 2018RvMPP...2....1H}. \\

The growth rate of the K-H instability highly influences by the presence and orientation of the magnetic field. If the uniform magnetic field is parallel to the wave vector and shear flow, it stabilizes the instability. The velocity shear component balances the magnetic tension component of the Lorentz force. However, when the magnetic field component is perpendicular to the shear flow, it does not affect the stability but modifies the growth rate of the K-H instability. In the present observation, the spine-associated magnetic field is similar to the thin magnetic flux tubes group in the fan-spine topology, and the two layers of plasma, "L1" and "L2" flow in this multilayer topology. We notice that the theoretical growth rate of the K-H instability possesses all the estimated parameters. For simplicity, we assume the condition where the parallel component of the magnetic field is parallel to the wave vector (i.e., $\vec{k}$.$\vec{B}$=kB). In the presence of the magnetic field and ignoring the gravity-driven Rayleigh-Taylor instability, the onset condition fo the Kelvin-Helmholtz instability is given as \citep{1961hhs..book.....C, 1998pfp..book.....C, 2017ApJ...850...60B, 2018NatSR...8.8136L, 2021JGRA..12629097S},

\begin{equation}
(\vec{k}.\Delta\vec{u})^{2} \ge (\rho_{1}+\rho_{2}){\frac{(\vec{k}.\vec{B_{1}})^{2}+(\vec{k}.\vec{B_{2}})^{2}}{\mu_{0}\rho_{1}\rho_{2}}}
\end{equation}
Here, $\vec{B}_{1}$ and $\vec{B}_{2}$ are the magnetic field vectors in two different layers L1 and L2. In the fan-spine topology, the elongated spine consists of a symmetric and uniform magnetic field. For simplicity, we assume that the magnitude of the magnetic field in the two layers of plasma (L1 and L2) is same, i.e., B$_{1}$=B$_{2}$. The most unstable mode of the K-H instability defines critical wave vector ($\vec{k}_{c}$), when $\vec{k}$$<$$\vec{k}_{c}$. The critical wave vector $\vec{k}_{c}$ is related to the critical wavelength ($\lambda_{c}$) of K-H instability. We conjecture the horizontal component of the magnetic as the critical magnetic field required to suppress the instability is \citep{2019ApJ...884L..51Y},
\begin{equation}
\lvert B_{c} \rvert \approx \frac{\lvert \Delta \vec{u} \rvert \sqrt{\mu_{0}\rho_{1}\rho_{2}}}{\sqrt{2(\rho_{1}+\rho_{2})}}
\end{equation}
We observe that the K-H unstable vortex appears in the lower layer of the plasma flow having mass density ($\rho_{1}$=7.6$\times$10$^{-15}$ g cm$^{-3}$). Therefore, we obtain the Alfv\'en speed in the second layer of plasma flow (L2), which consist of K-H unstable vortices as:
\begin{equation}
{V}_{A,2}=\frac{{\lvert \Delta \vec{u} \rvert}}{\sqrt{2(1+\frac{\rho_{1}}{\rho_{2}})}}
\end{equation} 
\begin{figure*}
\includegraphics[scale=1.0,angle=0,width=18.0cm,height=18.0cm,keepaspectratio]{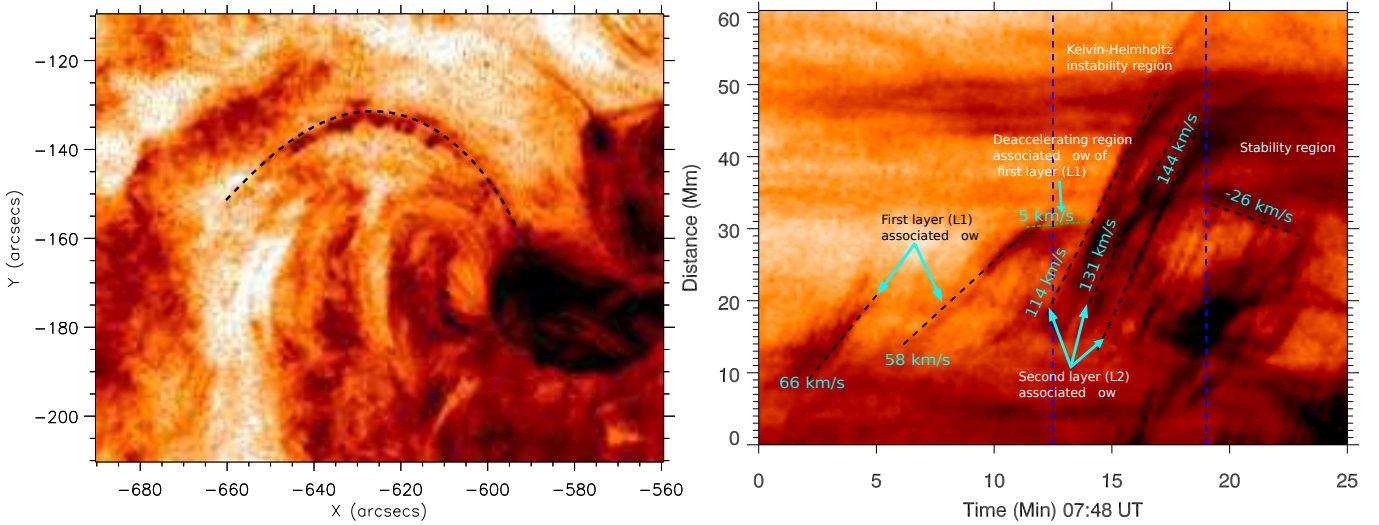}
\caption{Left panel: We select the slit along the interface of the two plasma flow to estimate the velocity of the ejected plasma. Right panel: We track the path of ejected plasma in the H–T diagram to measure the velocity of the K-H unstable vortices, velocity of the upper layer, and shearing or velocity difference between two layers.}
\end{figure*}

The estimated velocity difference of the two layers of the plasma flow is lying between $\approx$109--139 km s$^{-1}$ (Figure~5). Here we estimate the density ratio of K-H unstable plasma layer ($\rho_{1}$; Figure~6) and overlying plasma flow layer ($\rho_{2}$) is $\frac{\rho_{1}}{\rho_{2}}$=1.25. Therefore, the estimated Alfv\'en speed in the second layer of plasma flow ($V_{A,2}$=$\frac{\lvert\Delta \vec{u} \rvert}{2.1}$) is lying between 51--66 km s$^{-1}$. The estimated  Alfv\'en speed is consistent with the previously reported value in the cool and dense plasma \citep{2012A&A...540L..10I,  2019ApJ...874...57M}. A necessary condition for the onset of the K-H instability in the magnetic field-aligned flow is that the velocity difference between two layers must exceed twice of the minimum Alfv\'en speed \citep{2001RSPSA.457.1365H, 2019PhPl...26h2902H}. We found that the maximum and minimum velocity difference is greater than twice the minimum Alfv\'en speed (shearing velocity is 2.1 times the estimated Alfv\'en speed). We conjecture that the estimated parameter satisfies the standard criteria of the K-H instability i.e., $\lvert\vec{u}_{1}\rvert$ $>$ ${V}_{A,2}$ \citep{2004psci.book.....A}.\\

As abbreviated above, we observed a fan-spine configuration associated with an active region (AR 12297) on 2015 March 10 by using multiwavelength imaging observation of \textit{SDO}/AIA. In this paper, we investigate the evolution of K-H instability and then the stability of K-H type vortices that arise due to the twisting and shearing motions of the spine above the magnetic null point in the fan-spine configuration. For incompressible plasma, in the presence of sheared magnetic field and velocity field component (fan-spine topology), a transition occurs between K-H instability and tearing-mode instability when the parametric constant \citep{1986PhFl...29.2563E, 1997PhPl....4.1213D, 2013PhPl...20c2117W}, is given as follows:

\begin{equation}
\Lambda=\Bigg(\frac{L_{a}}{L_{b}}\Bigg)\Bigg({\frac{\lvert\Delta \vec{u}\rvert}{{V}_{A}}}\Bigg)^{{\frac{2}{3}}}
\end{equation}
Where $L_{a}$ and $L_{b}$ are the widths of the magnetic and velocity sheared layers, $\Delta$u is the shearing velocity or velocity difference between two layers, and $V_{A}$ is Alfv\'en speed far from the layer.  When $\Lambda \geq$1, the sheared velocity field is dominant, and the linear phase of the K-H instability regime evolves. However, when $\Lambda$ $<$1, the magnetic shear component is dominant over the velocity shear and causes the tearing mode instability. In the present work, we observe a fan-spine topology that appeared in the multi-wavebands of the \textit{SDO}/AIA. Figure~1 shows the fan-spine topology associated with an active region (AR 12297) in multiple EUV wavebands. The fan separatrix plane possesses both velocity and magnetic shear. Such a plane is known as the current-vortex sheet. \citet{2013PhPl...20c2117W} have simulated for the first time to detect the K-H instability evolved in the 3D current vortex sheet at the fan plane of a 3D null point. In the present study, multiple vortex-like structures develop at the interface of these two layers at a regular interval. A remarkable kink-like displacement appears in the lower panel (L2) of plasma flow as the interaction of the two layers is initiated at around 08:02 UT (Figures~2--3, 7). The difference between the linear phase of K-H instability and stable wave depends upon the growth rate of the wave amplitude. If the growth of wave amplitude is zero, then the wave is stable, but when the wave amplitude grows, it defines the linear phase of K-H instability \citep{2019ApJ...884L..51Y}. The linear K-H instability phase's characteristic wavelength ($\lambda$) defines the separation between two consecutive vortices. The observed vortices are lying in a curvilinear path with various intervals. Therefore, we measure the separation between the two consecutive vortices by using the equation $\lambda$=$\sqrt{(x_{2}-x_{1})^{2}+(y_{2}-y_{1})^{2}}$. This separation ($\lambda$) defines the characteristic wavelength of K-H instability (Figure~7). The separation between two consecutive vortices defines as the average value of the separation measured between either tip to tip, mid to mid, or tail to a tail distance of the two continuous K-H unstable vortex (Figure~7). \\

\begin{figure*}
\includegraphics[scale=1.0,angle=0,width=18.0cm,height=18.0cm,keepaspectratio]{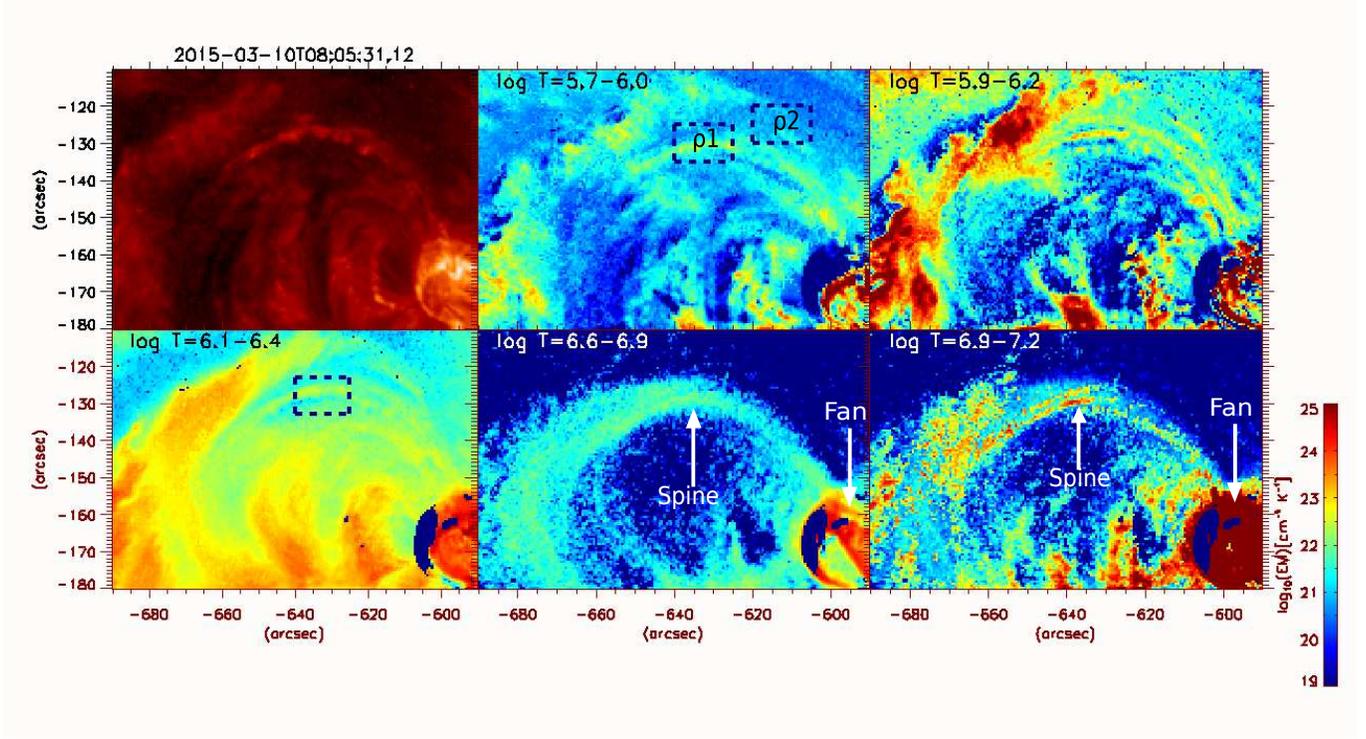}
\caption{DEM map at different temperatures of log $T$=5.7\,--\,7.2 at 08:05:32 UT when the K-H unstable vortices fully develop. Multi-thermal and multi-layer of plasma flow are evident in the fan-spine configuration. The K-H unstable vortices and flow inside the spine are primarily visible in coronal temperature ($\approx$log $T$=5.7\,--\,6.4), and fan-spine topology is evident in the hot channel (log $T$=6.8\,--\,7.2). We select two box regions to deduce the mass density in the two layers of plasma flow.}

\end{figure*}
 The estimated wavelength of K-H instability is lying $\approx$(4.9, 5.1, 5.5, 7.4)$\pm$1.1 $Mm$ along the curvilinear length of the spine (Figure~7, cyan color dotted connecting line). The uncertainty in the characteristic wavelength is equal to 1.1 Mm, which is equivalent to the spatial resolution of \textit{SDO}/AIA (1.5 arcseconds). In the present observation, the wavelength is growing along the elongated spine (Figure~7). The size of the K-H unstable vortices also varies from 2.2 Mm to 3.2 Mm. The thickness of these vortex-like structures (Figure~7; sawtooth pattern) lies between $\approx$2.2--3.2$\pm$1.1 Mm. The varying size of the vortex confirms that the wave amplitude is growing with the evolution of the instability. The estimated wavelength of K-H instability is $\approx$($\lambda$=4.9, 5.1, 5.5, 7.4)$\pm$1.1 $Mm$ lying between 4.9--7.4 Mm and half of the flow width as the estimated boundary layer thickness is (a=1.1--1.6 Mm; half of the wave amplitude) lying 1.1--1.6 Mm. \citet{1982JGR....87.7431M} performed a numerical simulation for the finite shear thickness layer. They found that the fastest-growing mode excited by the K-H instability should have a range. The linear growth of the fastest mode is a phase where the growth of the K-H instability is fastest. This mode is dominated by the early plasma dynamics of the K-H instability \citep{1987JGR....92.3195M}. The condition of the fastest-growing K-H mode is given by \citep{1982JGR....87.7431M, 2018NatSR...8.8136L}.\\
 
 \begin{equation}
 \lambda=(2-4)\times\pi\times a
 \end{equation}
  After putting the value of $\lambda$ and ''a'', it satisfied the condition as mentioned above, another verification for the evolution of K-H instability. \\
  
 The criteria for the K-H instability in the presence of sheared magnetic and velocity layer from Equation~8 is $\Lambda$ $\ge$1. To satisfy this criteria, we need to conjecture the Alfv\'en speed in the lower layer. The simultaneous presence of the sheared velocity layer and magnetic layer in the fan-spine configuration shows by using the composite image of \textit{SDO}/AIA 94+304 {\AA} wavebands (Figure~8). We selected a path perpendicular to the spine at maximum height to measure the width of the magnetic layer (top right panel of Figure~8; appeared in hot plasma in AIA 94 {\AA} wavelength) and sheared velocity layer that consists of the K-H unstable vortices (bottom right panel of Figure~8; appeared in the cool AIA 304 {\AA} filter). As discussed above, the formation and propagation of these vortices are very rapid ($\approx$114--144 km s$^{-1}$). Therefore, we manually track the maximum width of the thickest vortex by using the cursor command. The magnetic layer and sheared velocity layer's estimated width is 3.50$\pm$1.1 Mm and 9.0$\pm$1.1 Mm. From Equation~8, the criteria for the evolution of K-H instability in the sheared magnetic and velocity layer is $\Lambda$ $\ge$ 1. The estimated width of the magnetic layer (L$_{a}$) and sheared velocity layer ($L_{b}$) is $\approx$9 and $\approx$3.5 $Mm$. Therefore, the parametric measurement $\Lambda\ge$1 condition is also satisfied for the evolution of the K-H instability.\\ 	                                                                                                                                                               
\begin{figure*}
\includegraphics[scale=1.0,angle=90,width=18.0cm,height=18.0cm,keepaspectratio]{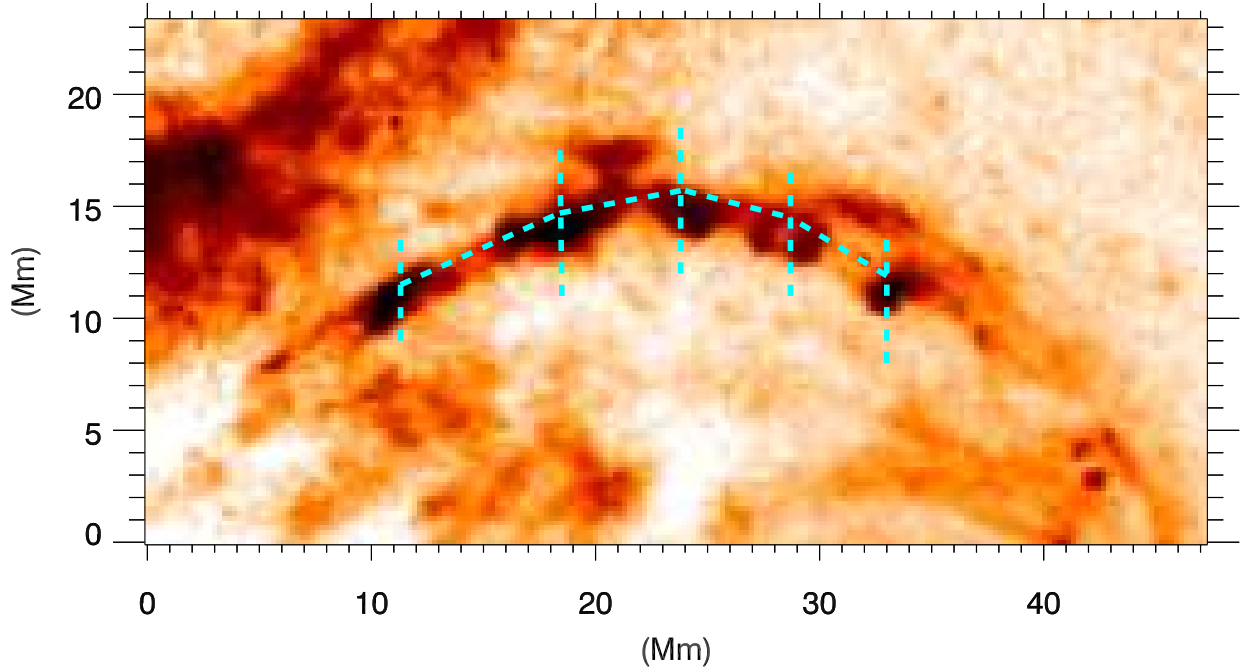}
\caption{The region of interest (ROI; white box in Figure 5) is displayed using \textit{SDO}/AIA 304 {\AA} images at 08:05:32 UT on 2015 March 10 when the K-H unstable vortices are completely evolved. The reverse color contrast is used to understand the evolution of sheared plasma vortices as well as the launch of the magnetic twists at different epochs.}
\end{figure*}

\section{Discussion and Conclusion}
We have analyzed an active region (AR 12279) observed on 2015 March 10 by using multiwavelength EUV wavebands of \textit{SDO}/AIA, which possesses a fan-spine structure. The observed fan-spine configuration consists of both hot and cool plasma simultaneously as observed by multiwavelength imaging observation from \textit{SDO}/AIA. This work demonstrates the evolution, formation of K-H unstable vortices, and stability of K-H instability in a fan-spine configuration. This configuration consists of the sheared magnetic component and the velocity layer known as the current-vortex sheet. \citet{2013PhPl...20c2117W} have simulated for the first time to observe the Kelvin-Helmholtz instability in a 3-dimensional current-vortex sheet at the fan plane of a 3D magnetic null point. However, the present work is the first observational effort to understand the evolution of K-H instability and its instability inside an elongated hot spine. The shearing motion around the outer spines may be the cause for the growth of this instability.  \\

The uniqueness of the present observation is that the multi-thermal plasma (Figures~1, 6, 8; 0.5--3 MK) started to lift up from the hot fan plane (Figures~1, 6; $\approx$10 MK) and interacted within the ambient hot elongated spine (Figures~;1, 6$\approx$5--8 MK). The multiwavelength imaging observation of \textit{SDO}/AIA (Figure~1; upper panel) and thermal analysis confirm that the hot fan-spine structure is already present. The first layer of flow (L1) consists of multiple plasma blobs shot along the length of the elongated spine (Figures~2--6). After reaching the maximum height, this layer initiates to decelerate along the same path. The second layer of cool plasma (Figures~2--3; L2) just below the first layer started to lift and interact inside the hot spine. The upper layer (L1) lift with a velocity of $\approx$58-66 km s$^{-1}$ before decelerating to 5 km s$^{-1}$, whereas multiple vortex-like plasma blobs in the lower layer (L2) propagate with different velocities in the range 114--144 km s$^{-1}$ (Figure~5). The second layer moves with greater velocity and causes the onset of Kelvin-Helmholtz instability in the hot spine. We have also estimated the separation between two consecutive vortexes (known as characteristic wavelength) and found that it increases, satisfying the fastest-growing K-H mode. We have measured the differential emission measure (DEM) using six AIA filters AIA 94 (log T=6.8), AIA 131 (log T = 5.6, 7.0), AIA 171 (log T = 5.8), AIA 193 (log T = 6.2), AIA 211 (log T = 6.3), and AIA 335 (log T = 6.4) to understand the thermal properties of fan-spine topology, evolution of K-H instability, and their stability phase. It is useful to measure the mass density within the K-H unstable vortex and overlying regions to deduce K-H instability's growth rate. We have also found that the K-H unstable vortices are associated with the cool plasma (0.5--3 MK; Figure~6). However, as the lower layer (L2) gets stable, this layer's boundary becomes smooth, and the local plasma has heated up to the $\approx$8 MK (Figure~6). The DEM diagnostics show that the lower layer (L2) is associated with denser plasma ($\rho_{2}$) as compared to the overlying first layer of plasma flow (Figure~6). We have estimated shearing velocity or velocity difference lying between 109--139 km s$^{-1}$ (Figure~5). In the presence of the magnetic field, the growth rate of the K-H instability significantly modifies. We deduce the Alfv\'en speed by using Equation~7 in the second layer of plasma flow. We found that the Alfv\'en speed in the second layer of plasma flow is lower than the velocity difference and velocity in the lower layer. We conjecture that the calculated Alfv\'en speed satisfy the standard criteria of the K-H instability i.e., $\lvert \vec{u}_{1}\rvert$ $>$ $V_{A,2}$ \citep{1978SoPh...58...57P, 2004psci.book.....A}. The estimated velocity in the lower layer and velocity difference is higher than the estimated Alfv\'en speed, which also confirms the evolution of K-H instability inside the hot spine configuration. We have also performed a measure of the parametric constant $\Big($ $\Lambda$ = ($\frac{L_{a}}{L_{b}})(\frac{\lvert\Delta \vec{u}\rvert}{V_{A}})^{\frac{2}{3}}\Big)$ and found that $\Lambda\ge$1. In the presence of magnetic and sheared velocity layers, when $\Lambda\ge$1, the velocity shear dominates the layer, and the linear phase of the K-H instability develops \citep{1986PhFl...29.2563E, 2013PhPl...20c2117W}. Our result is also consistent with this measured parametric constant. Therefore, we conclude that the K-H instability evolves inside the hot fan-spine configuration. Two different flow layers propagate with different velocities in the opposite direction, and their interaction may cause the onset of the K-H instability in the hot spine. The fragmentation via the K-H instability inside the spine provides a rapid mechanism to circulate the mass and energy locally in the solar atmosphere.\\
\begin{figure*}
\includegraphics[scale=1.0,angle=0,width=18.0cm,height=18.0cm,keepaspectratio]{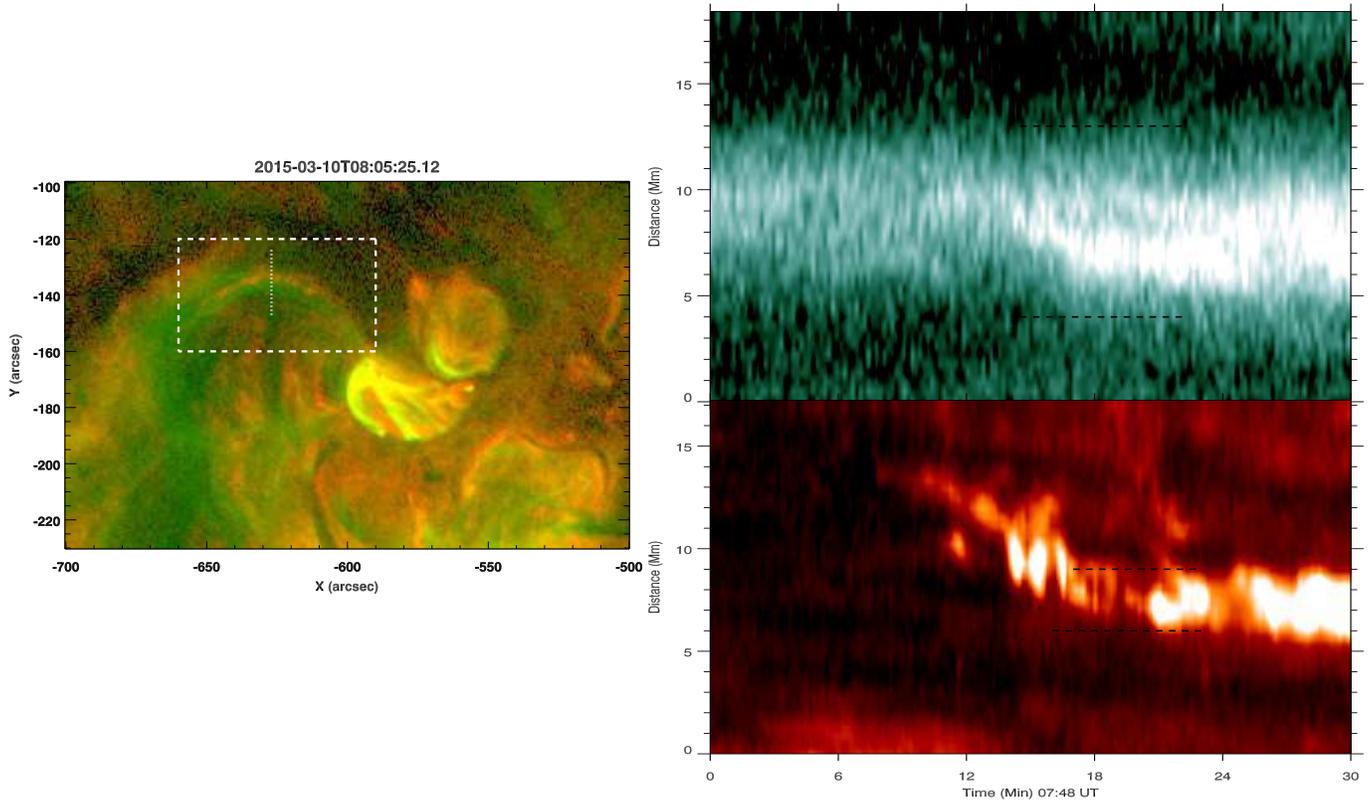}
\caption{Left panel: The composite image of \textit{SDO}/AIA 94+304 {\AA} shows the simultaneous presence of the hot and cool plasma in the fan-spine configuration. The hot loops are already present in the ambient corona. The K-H unstable cool plasma traps within the outer spine of the fan-spine topology. A vertical path has been selected along the width of the spine to estimate the approximate width of the magnetic and velocity sheared layer. Right panel: Hight-time diagram along the width of the spine.}
\end{figure*}

To the best of our knowledge, the present work provides the first observational evidence of the development of K-H instability in the fan-spine configuration. It also provides a rapid mechanism to transfer the mass and energy between two distinct regions separated by the fan-spine configuration. We find that these instability features develop on different spatio-temporal scales inside the hot elongated spine. Therefore, we provide an overall scenario for the evolution of the linear phase of K-H instability and its stability phase in the fan-spine configuration. In conclusion, the present study provides observational evidence of K-H instability's morphological evolution in the fan-spine topology. Further observations are required to understand the rapid mass transfer in the localized region of the solar atmosphere. The upcoming observations from the Aditya-L1 will be useful to understand such multi-thermal plasma dynamics in the solar atmosphere using SUIT and VELC data up to mid-corona.
\section*{Acknowledgments}
We acknowledge the constructive comments of referee that improved the manuscript. S.K. Mishra is thankful to Professor S.P. Rajaguru for discussion and his fruitful suggestions. A.K. Srivastava acknowledges the UKIERI Research Grant for the support of his research. S.K. Mishra acknowledges the Department of physics, Indian Institute of Technology for providing him institute fellowship and computational facilities. B. Singh would like to acknowledge the Council of Scientific \& Industrial Research (CSIR), Government of India, for financial support through a Senior Research Fellowship (CSIR-SRF). We acknowledge the use of \citep{2015ApJ...807..143C} for calculating the differential emission measure (DEM). Data courtesy of NASA/SDO and the AIA science team.

\end{document}